# Polarization, transverse shifts, and angular momentum conservation laws in partial reflection and refraction of an electromagnetic wave packet

K.Yu. Bliokh[1,2*] and Yu.P. Bliokh[3]

[1]*Institute of Radio Astronomy, 4 Krasnoznamyonnaya St., Kharkov, 61002, Ukraine*
[2]*Amcrys Ltd., 60 Lenin Ave., Kharkov, 61001, Ukraine*
[3]*Department of Physics, Technion, Haifa, 32000, Israel*

We present a solution to the problem of partial reflection and refraction of a polarized paraxial Gaussian beam at the interface between two transparent media. The Fedorov–Imbert transverse shifts of the centers of gravity of the reflected and refracted beams are calculated. Our results differ in general case from those derived previously by other authors. In particular, they obey general conservation law for the beams' total angular momentum but do not obey one-particle conservation laws for individual photons, which have been proposed by Onoda *et al.*, [Phys. Rev. Lett. **93**, 083901 (2004)]. We ascertain that these circumstances relate to the accepted in the literature artificial model for the polarized beam which does not fit to real beams. The present paper resolves the recent controversy and confirms the results of our previous paper [Bliokh *et al.*, Phys. Rev. Lett. **96**, 073903 (2006)]. In addition, a diffraction effect of angular transverse shifts of the reflected and refracted beams is described.

PACS numbers: 41.85.-p, 42.15.-i, 42.25.Gy, 42.25.Ja

## I. INTRODUCTION

Reflection and refraction of a polarized plane electromagnetic wave at the interface between two homogeneous isotropic media is described by the Snell law and Fresnel formulas [1]. However, real confined fields – wave packets or beams – consist of an infinite set of plane waves with different wave vectors. Clearly each of the waves satisfies the Snell and Fresnel laws, but their superposition can behave in an uncommon way. In particular, depending on polarization of the incident beam, the center of gravity of the reflected or refracted beam undergoes a *transverse shift* (TS) and leaves the plane of incidence. This is the Fedorov–Imbert shift (or the lateral shift) which has been considered in a number of theoretical and experimental papers [2–11]. In spite of the small magnitude (of the order of the wavelength), TS has a fundamental meaning: it provides for the conservation of the *total angular momentum* (TAM) of the beams, including intrinsic, spin part [4,5,7–10,12]. Furthermore, TS allows to observe the spin-Hall effect of photons: the splitting of a linearly-polarized beam into two circularly polarized ones at reflection or refraction [3,8]. TS also occurs in the reflection or refraction of elastic waves [10] and of beams of arbitrary nature with vortices [11–13]. In the latter case the effect can be noticeably enhanced owing to a large value of the intrinsic angular momentum carried by the beam.

Although numerous investigations have been carried out, the issue of the final formulation of the TAM conservation law and expressions for TS is still open. Fedoseev [5] has derived the general expressions for TSs of the reflected and refracted beams valid for any paraxial incident beam. These papers, however, do not contain explicit TS values for a particular (for instance, Gaussian) beam. Recent publications caused a controversy in the literature. On the one hand, Onoda *et al.* [7,9] propose an approach in which TSs of the reflected and refracted wave packets obey two TAM conservation laws for individual photons. They argue that each photon in the incident packet can be either reflected or refracted and respective one-photon conservation laws control these processes. Detailed analytic calculations of TSs for partial reflection of a wave packet in the paraxial approximation (which are equivalent to calculations of earlier paper [6]) as well as numerical simulations confirm their results. On the other hand, in recent paper [8] we have also derived analytic expressions for fields of the reflected and refracted paraxial Gaussian beams. Our results indicate that TSs of the reflected and refracted beams satisfy the general TAM conservation law for beams, but in generic case do not obey the one-photon conservation laws. (A fundamental reason for that we see in the interference and lack of the "which path" information in two-channel wave scattering, which cannot be explored through one-particle considerations [8].)

---
[*]E-mail: k_bliokh@mail.ru

In the present paper we resolve a discrepancy between the results of papers [6,7,9] and [8]. We show that the results differ because of the distinction in models for the incident polarized beam. The beams dealt with in papers [6,7,9] possesses an artificial polarization structure which facilitates the calculations but cannot correspond to a real polarized beam. For instance, a pure linearly-polarized beam cannot be constructed in that model and the incident beam's field in the accompanying coordinate system surprisingly depends on the angle of incidence. On the contrary, a more detailed analysis corroborates our prior results and validity of our model of the polarized beam.

## II. BASIC RELATIONS AND CONSERVATION LAWS

We will deal with polarized monochromatic electromagnetic beams. The beam supposed to be semiclassical, i.e. its characteristic dimensions are large as compared to the wavelength, which enables one to use the paraxial approximation and to talk about the beam polarization. In this way, the beam wave front represents near-plane wave with practically uniform polarization. (In fact, the plane waves constituting the beam propagate at different small angles and have slightly different polarizations. This circumstance plays a crucial role and will be discussed below.)

Let us consider the partial reflection and refraction of a polarized monochromatic beam of frequency $\omega$ at an interface between two homogeneous isotropic lossless media with refractive indices $n_1 = \sqrt{\varepsilon_1 \mu_1}$ and $n_2 = \sqrt{\varepsilon_2 \mu_2}$ ($\varepsilon_i$ and $\mu_i$, $i = 1, 2$, are the permittivities and permeabilities of the media), Fig. 1. We will also use relative characteristics $n = n_2 / n_1$, $\varepsilon = \varepsilon_2 / \varepsilon_1$, $\mu = \mu_2 / \mu_1$. Along with standard coordinate system $(x, y, z)$, attached to the interface $z = 0$ and incidence plane $y = 0$, we introduce beam coordinate systems, $(X^a, Y^a, Z^a)$. In what follows $a = (i), (r), (t)$, and superscripts $(i)$, $(r)$, and $(t)$ point at relation to the incident, reflected, and refracted beams, respectively (Fig. 1). In $(X^a, Y^a, Z^a)$ coordinate system $Y^a$ axis coincides with $y$ axis, whereas $Z^a$ axis is directed along the wave vector of the beam's central plane wave, $\mathbf{k}_c^a$. The origins of all coordinate systems are located at the scattering point, defined as an intersection of the incident packet center and the interface $z = 0$. Angles of propagation between wave vectors $\mathbf{k}_c^a$ and $z$ axis are denoted as $\vartheta^a$. We will also use notations (Fig. 1)

$$\theta = \vartheta^{(i)} = \pi - \vartheta^{(r)}, \quad \theta' = \vartheta^{(t)}. \tag{1}$$

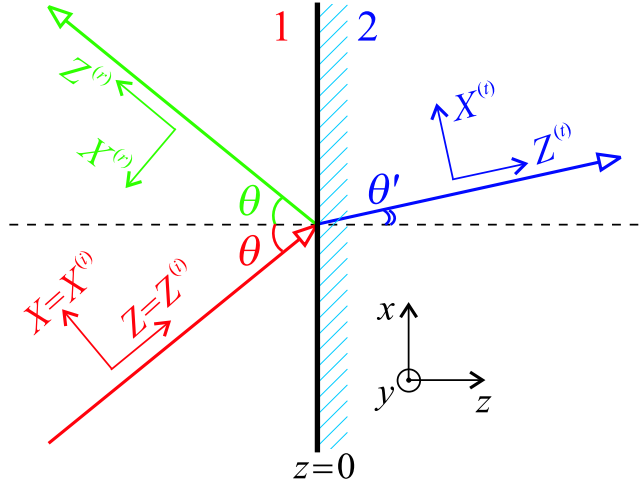

**Fig. 1.** (Color online). The scheme of the wave reflection and refraction with beam coordinates used in the text.

Various conservation laws govern the beam scattering at the interface. They can be formulated if one imagines that we deal with a localized wave packet rather than with an infinite beam. The packet length can be arbitrarily large to make its spectrum arbitrarily narrow in order to eliminate the distinction between the packet and monochromatic beam with frequency $\omega$. Let $a$ th packet include $N^a$ photons, i.e. its field energy is $W^a = N^a \omega^a$ (we use units $\hbar = c = 1$).

First, time invariance of the problem leads to the constancy of frequencies in all the packets,
$$\omega^a = \omega = const, \qquad (2)$$
and provides for conservation of the total energy, $W^{(i)} = W^{(r)} + W^{(t)}$, or, in view of Eq. (2), of the total number of photons in the scattering process:
$$N^{(i)} = N^{(r)} + N^{(t)}. \qquad (3)$$
By introducing the energy reflection and refraction coefficients $\mathcal{R} = W^{(r)}/W^{(i)} = N^{(r)}/N^{(i)}$ and $\mathcal{T} = W^{(t)}/W^{(i)} = N^{(t)}/N^{(i)}$, Eq. (3) can be written as
$$\mathcal{R} + \mathcal{T} = 1, \qquad (4)$$
Note, that the total field energy in $a$ th packet can be estimated as $W^a \propto \varepsilon^a \left|\mathbf{E}_c^a\right|^2 V^a$ (with $\varepsilon^{(i)} = \varepsilon^{(r)} \equiv \varepsilon_1$ and $\varepsilon^{(t)} \equiv \varepsilon_2$), where $\mathbf{E}_c^a$ is the electric field in the center of the wave packet, $V^a$ is the volume of packet, and the volume changes at the refraction as $V^a = V n_1 \left|\cos\vartheta^a\right| / n^a \cos\theta$ ($V \equiv V^{(i)}$). Then $\mathcal{R} \simeq \mathcal{R}_c$, $\mathcal{T} \simeq \mathcal{T}_c$, (where $\mathcal{R}_c$ and $\mathcal{T}_c$ are the energy coefficients for central plane waves in packet), and $\mathcal{R}_c = R_c^2$ and $\mathcal{T}_c = \dfrac{n\cos\theta'}{\mu\cos\theta}T_c^2$, where $R_c, T_c = \left|\mathbf{E}_c^{(r,t)}\right|/\left|\mathbf{E}_c^{(i)}\right|$ are the amplitude (Fresnel) reflection and refraction coefficients for the central plane waves. As a result, the conservation law (4) can be represented in the form known for the amplitude coefficients:
$$R_c^2 + \frac{n\cos\theta'}{\mu\cos\theta}T_c^2 = 1. \qquad (5)$$

Second, the translation invariance of the problem along $x$ and $y$ axes brings about the conservation of the respective wave vectors components:
$$k_{cx}^a = const, \quad k_{cy}^a = const = 0. \qquad (6)$$
This provides conservation of the corresponding components of the total momentum. The momentum of the $a$ th packet is $\mathbf{p}^a \simeq N^a \mathbf{k}_c^a$, and the conservation law for the total momentum, $p_{x,y}^{(i)} = p_{x,y}^{(r)} + p_{x,y}^{(t)}$, with Eq. (6), also leads to the energy (photon number) conservation,

Eqs. (3)–(5). The second equation (6) means that all vectors $\mathbf{k}_c^a$ are parallel to $y=0$ plane, whereas the first equation (6) together with the dispersion law $\omega = k^a = n^a k_0$ ($n^{(i)} = n^{(r)} \equiv n_1$, $n^{(t)} \equiv n_2$, $k^a = k_c^a$, and $k_0$ is the wave vector in vacuum) results in the Snell law:

$$n^a \sin \vartheta^a = const, \text{ or } \sin\theta = n\sin\theta'. \qquad (7)$$

Note that equations (2) and (6) can be regarded as conservation laws for energy and tangent momentum components of a single, reflected or refracted, photon. For instance, the energy conservation for one photon gives $\omega^{(i)} = \omega^{(r)}$ at the reflection and $\omega^{(i)} = \omega^{(t)}$ at the refraction, which together implies Eq. (2).

Finally, the axial symmetry of the problem with respect to $z$ axis results in the conservation of $z$ component of the total angular momentum in the problem. The TAM density (TAM per one photon), $\mathbf{j}^a$, can be represented as a sum of the orbital angular momentum and intrinsic (or spin) angular momentum [7–9]:

$$\mathbf{j}^a \simeq \mathbf{r}_c^a \times \mathbf{k}_c^a + \sigma_c^a \mathbf{k}_c^a / k^a. \qquad (8)$$

Here $\mathbf{r}_c^a$ is the radius-vector of the wave packet's center of gravity and $\sigma_c^a \in [-1,1]$ is the mean helicity of the packet, i.e. the difference between numbers of right-hand and left-hand photons divided by the total number of photons. If one introduces a two-component complex unit vector of the polarization of the wave packet center, $|\zeta_c^a) = \begin{pmatrix} \zeta_c^{+a} \\ \zeta_c^{-a} \end{pmatrix}$, represented in the basis of circular polarizations (helicity basis), then

$$\sigma_c^a = (\zeta_c^a | \hat{\sigma}_3 | \zeta_c^a) = |\zeta_c^{+a}|^2 - |\zeta_c^{-a}|^2, \qquad (9)$$

where $\hat{\sigma}_3 = \text{diag}(1,-1)$ is the Pauli matrix. Taking the geometry of the problem into account, the $z$ component of TAM per photon, Eq. (8), equals

$$j_z^a = -y_c^a k_c^a \sin\vartheta^a + \sigma_c^a \cos\vartheta^a. \qquad (10)$$

It may seem that, by an analogy with the one-photon energy and momentum conservation laws (2) and (6), one-photon conservation laws for $z$ component of TAM, Eq. (10), should also be valid:

$$j_z^a = const, \text{ or } j_z^{(i)} = j_z^{(r)}, \ j_z^{(i)} = j_z^{(t)}, \qquad (11)$$

Conservation laws (11) have been proposed in papers [7,9]. However, in our recent paper [8] and below we show that in general equations (11) are not satisfied for polarized Gaussian beams. The principal distinction of conservation laws (11) as compared to Eqs. (2) and (6) is as follows. Conservation laws for the energy and momentum, Eqs. (2) and (6), are written in the zero approximation when the wave packet is replaced by its central plane wave. At the same time, TAM of a polarized plane wave vanishes and all non-trivial effects related to non-zero TAM originate from the confinement of the packet or beam in the transverse direction (see, for instance, [14]). Therefore, when involving a non-zero TAM, one should remember that we deal with essentially localized packet which consists of a set of interfering plane waves with different wave vectors. In general case this does not validate the conservation law with parameters of the single central plane wave. Paper [7] argued that the partial reflection and refraction process consists of one-photon acts of pure reflection and refraction, and TAM conservation laws (11) must be fulfilled in each act. However, it is shown in [8] that one-photon considerations and conservation laws (11) evoke "which path" information in the process which represents two-channel wave scattering with interference. It is known from quantum mechanics that "which path" information eliminates the interference pattern and, therefore, equations (11) cannot be applied to the classical wave problem of partial reflection and refraction of a wave packet.

Nevertheless, the general TAM conservation law takes place in the problem. TAM of the wave packet equals $\mathbf{J}^a = N^a \mathbf{j}^a$, and conservation law for $z$ component of TAM, $J_z^{(i)} = J_z^{(r)} + J_z^{(t)}$, takes the following form:

$$j_z^{(i)} = \mathcal{R}_c j_z^{(r)} + \mathcal{T}_c j_z^{(t)} \quad \text{or} \quad j_z^{(i)} = R_c^2 j_z^{(r)} + \frac{n \cos \theta'}{\mu \cos \theta} T_c^2 j_z^{(t)}. \tag{12}$$

If conservation laws (11) hold true, then Eq. (12) is reduced to the same energy (photon number) conservation law (3)–(5) [9], but in the absence of Eqs. (11) equation (12) is an independent conservation law in the problem.

To express equation (12) in terms of characteristics of the incident packet, two media, and unknown transverse shifts, one should find reflection and refraction coefficients as well as polarization characteristics of the reflected and refracted waves determined from the Fresnel equations for central plane waves of the packets. Polarization in the center of the wave packet can be represented as $\mathbf{e}_c^a = \dfrac{\mathbf{u}_{X^a} + m_c^a \mathbf{u}_y}{\sqrt{1 + |m_c^a|^2}}$, where $\mathbf{u}_\alpha$ are the unit vectors along the corresponding axes and $m_c^a$ is a complex parameter characterizing polarization of the wave packet center in the basis of waves linearly polarized along $X^a$ and $y$. Polarization vector $|\zeta_c^a)$ introduced above and helicity $\sigma_c^a$ are expressed via $m_c^a$ as

$$|\zeta_c^a) = \begin{pmatrix} 1 - i m_c^a \\ 1 + i m_c^a \end{pmatrix} \Big/ \sqrt{2\left(1 + |m_c^a|^2\right)}, \quad \sigma_c^a = \frac{2 \operatorname{Im} m_c^a}{1 + |m_c^a|^2}. \tag{13}$$

Quantities $R_c$, $T_c$, $m_c^{(r)}$, and $m_c^{(t)}$ are the functions of polarization of the incident wave, $m_c^{(i)} \equiv m_c$, angle of incidence, $\theta$, and parameters of media. From the Fresnel equations [1] it follows that

$$R_c = \frac{\sqrt{|R_{\|c}|^2 + |R_{\perp c}|^2 |m_c|^2}}{\sqrt{1 + |m_c|^2}}, \quad T_c = \frac{\sqrt{|T_{\|c}|^2 + |T_{\perp c}|^2 |m_c|^2}}{\sqrt{1 + |m_c|^2}}, \quad m_c^a = \rho_c^a m_c. \tag{14}$$

Here $\rho_c^{(i,r,t)} = 1, R_{\perp c}/R_{\|c}, T_{\perp c}/T_{\|c}$, whereas $R_{\|,\perp c}$ and $T_{\|,\perp c}$ are the Fresnel reflection and refraction coefficients for central plane waves linearly polarized along $X^a$ and $y$ axes (i.e., with the electric field lying in the plane of incidence $y = 0$ and orthogonal to it, respectively) [1]:

$$T_{\|c} = \frac{2 n \cos \theta}{\varepsilon \cos \theta + n \cos \theta'}, \quad T_{\perp c} = \frac{2 \mu \cos \theta}{\mu \cos \theta + n \cos \theta'}, \quad R_{\|c} = 1 - \frac{\cos \theta'}{\cos \theta} T_{\|c}, \quad R_{\perp c} = T_{\perp c} - 1. \tag{15}$$

By substituting Eqs. (9), (10), (13), and (14) into Eq. (12) and introducing the transverse shifts of the beams' centers of gravity with respect to the incident beam's center, $\Delta^a = y_c^a - y_c^{(i)}$, we obtain

$$\mathcal{R}_c \Delta^{(r)} + \mathcal{T}_c \Delta^{(t)} = -\frac{2 \operatorname{Im} m_c}{k\left(1 + |m_c|^2\right)} \cot \theta \left[1 + R_{\|c} R_{\perp c} - \frac{n \cos^2 \theta'}{\mu \cos^2 \theta} T_{\|c} T_{\perp c}\right], \tag{16}$$

where $k \equiv k^{(i)}$.

Equation (16) indicates that at least one of the TSs, $\Delta^{(r)}$ or $\Delta^{(t)}$, is non-zero when $\operatorname{Im} m_c \neq 0$, i.e. in the case of non-planar (elliptical) polarization of the incident wave. It is impossible to determine two unknown values of $\Delta^{(r)}$ and $\Delta^{(t)}$ from one conservation law (16) – one has to solve a complete problem of partial reflection and refraction of a confined wave packet or beam. If one-photon TAM conservation laws (11) are valid, then two equations (11) laws enables one to find the values of TSs:

$$\Delta^a = \frac{2 \operatorname{Im} m_c}{k} \cot \theta \left[\frac{\rho_c^a \cos \vartheta^a / \cos \theta}{1 + \rho_c^{a2} |m_c|^2} - \frac{1}{1 + |m_c|^2}\right]. \tag{17}$$

These values have been obtained in papers [6,7,9] both from the conservation laws (11) and as a result of rigorous solution of the beam scattering problem. However, TSs found in [8], likewise through rigorous solution of the problem of partial reflection and refraction of a Gaussian beam, in generic case do not coincide with Eq. (17) and, as was noted above, fulfill conservation law (12), but not (11). The reason of this contradiction will be elucidated below.

### III. TWO MODELS FOR POLARIZED INCIDENT BEAM

When exploring rather heuristic and qualitative considerations in the previous Section, from hereon we will consider a complete electrodynamical problem of partial reflection and refraction of a paraxial polarized beam. Above all, we have to set the field of the incident beam.

It can be constructed in the wave vector representation. The beam consists of an infinite set of plane waves with the wave vectors close to the central one, $\mathbf{k}_c$. [All quantities in this Section are related to the incident beam and we omit superscripts (*i*) throughout the Section.] The wave vector of a partial plane wave from the incident beam can be represented as $\mathbf{k} = \mathbf{k}_c + \boldsymbol{\kappa}$, where $\boldsymbol{\kappa}$ is distributed around the zero vector and in paraxial, linear in $\mathbf{v} = \boldsymbol{\kappa}/k$, approximation one can reckon $\boldsymbol{\kappa} \perp \mathbf{k}_c$. Note that wave vector $\mathbf{k}$ of a partial plane wave, in general, does not belong to the plane of incidence $y = 0$ and the angle of incidence corresponding to it differs from $\theta$. To describe propagation of the current plane wave we introduce the respective coordinate system $(X', Y', Z')$ (analogous to $(X, y, Z)$ for the central wave, Fig. 1), similarly as it was done in [5–7,9], Fig. 2. In so doing, the $Z'$ axis is attached to the wave vector $\mathbf{k}$, whereas $X'$ axis lies in incident plane determined for the given plane wave. This coordinate system will be used exclusively for projections of the electric field vector, which are independent of position of its origin.

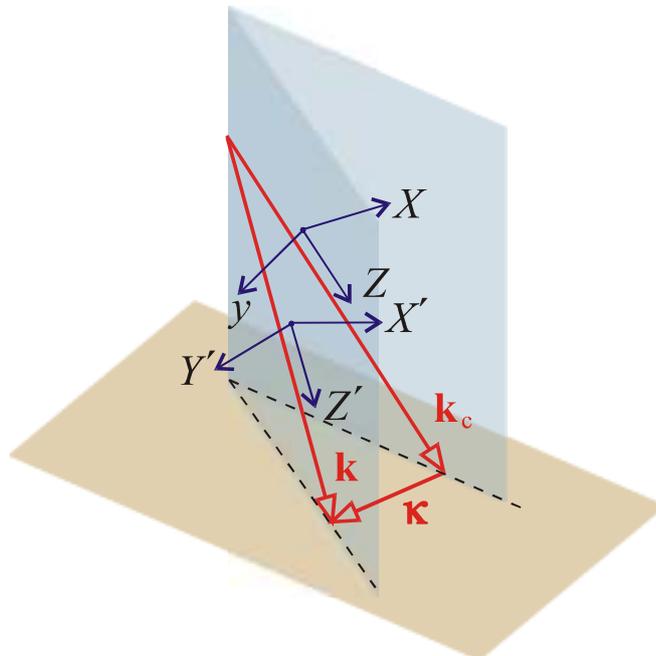

**Fig. 2.** (Color online). Central and non-central wave vectors in the incident beam, attendant coordinate frames and respective incident planes.

We should specify the polarization of each plane wave, $\tilde{\mathbf{e}} = \tilde{\mathbf{e}}(\boldsymbol{\kappa})$. [Here and in what follows the quantities in the wave vector representation are marked by tilde.] Polarization of the central wave have been defined above as $\mathbf{e}_c = \tilde{\mathbf{e}}_c \equiv \tilde{\mathbf{e}}(0) = \dfrac{\mathbf{u}_X + m_c \mathbf{u}_y}{\sqrt{1+|m_c|^2}}$; it is characterized by the complex number $m_c$. There are two ways to define polarizations of other waves.

The first one is to characterize polarization of each wave by the same number $m_c$ but in its *own* coordinate system $(X', Y', Z')$:

$$\tilde{\mathbf{e}}^{[I]} = \frac{\mathbf{u}_{X'} + m_c \mathbf{u}_{Y'}}{\sqrt{1+|m_c|^2}}. \tag{18}$$

This way of specifying the polarization was utilized in [6,7,9]. It can be shown that orts of coordinate systems $(X, y, Z)$ and $(X', Y', Z')$ in the approximation under consideration are connected by relations

$$\begin{aligned}
\mathbf{u}_{X'} &= \mathbf{u}_X + \nu_y \cot\theta\, \mathbf{u}_y - \nu_X \mathbf{u}_Z, \\
\mathbf{u}_{Y'} &= \mathbf{u}_y - \nu_y \cot\theta\, \mathbf{u}_X - \nu_y \mathbf{u}_Z, \\
\mathbf{u}_{Z'} &= \mathbf{u}_Z + \nu_X \mathbf{u}_X + \nu_y \mathbf{u}_y.
\end{aligned} \tag{19}$$

Hence, polarization vector (18) in the central coordinate system of the beam, $(X, y, Z)$, takes the form

$$\tilde{\mathbf{e}}^{[I]} = \frac{\left(1 - m_c \nu_y \cot\theta\right)\mathbf{u}_X + \left(m_c + \nu_y \cot\theta\right)\mathbf{u}_y - \left(\nu_X + m_c \nu_y\right)\mathbf{u}_Z}{\sqrt{1+|m_c|^2}}. \tag{20}$$

In this coordinate system a complex parameter $\tilde{m}^{[I]} \equiv \tilde{e}_X^{[I]} / \tilde{e}_y^{[I]}$ corresponds to the polarization (18) or (20):

$$\tilde{m}^{[I]} = \frac{m_c + \nu_y \cot\theta}{1 - m_c \nu_y \cot\theta}. \tag{21}$$

The second way to specify the polarization, which has been used in [8], is to consider polarization of all plane waves in the projection onto the *common* coordinate system $(X, y, Z)$, attached to the central wave. By requiring that $\tilde{e}_y^{[II]} / \tilde{e}_X^{[II]} = m_c$ and $\tilde{\mathbf{e}}^{[II]} \mathbf{k} = 0$, we get

$$\tilde{\mathbf{e}}^{[II]} = \frac{\mathbf{u}_X + m_c \mathbf{u}_y - \left(\nu_X + m_c \nu_y\right)\mathbf{u}_Z}{\sqrt{1+|m_c|^2}}. \tag{22}$$

Thus, in this model of the beam the polarization is uniform in the beam coordinate system:
$$\tilde{m}^{[II]} = m_c.$$

By analyzing equation (21), it can be shown that parameters $\tilde{m}^{[II]} = m_c$ and $\tilde{m}^{[I]}$ correspond to polarizations with the same eccentricity of the polarization ellipse but with different orientations of the ellipse. In particular, the circular polarization $m_c = \pm i$ gives rise to the same circular polarization $\tilde{m}^{[I]} = \pm i$. In the case of linear polarization $m_c$, $\operatorname{Im} m_c = 0$, the polarization $\tilde{m}^{[I]}$ will also be linear, $\operatorname{Im} \tilde{m}^{[I]} = 0$, but with different orientation. For instance, if $m_c = 0$ or $m_c = \infty$ (which corresponds to linear polarizations along $X$ or $y$, respectively), then $\tilde{m}^{[I]} = \nu_y \cot\theta$ or $\tilde{m}^{[I]} = -1/\nu_y \cot\theta$, i.e. polarization of a plane wave with $\kappa_y \neq 0$ turns out to be inclined with respect to $X$ or $y$ axes, respectively.

The above examples demonstrate that in actual fact the polarization structure of the beam of [6,7,9], Eqs. (18) and (20), has a strange and unnatural form in the attendant coordinate system of the beam, $(X, y, Z)$. Specifically, the polarization vector of the incident beam turns out to be dependent on the incidence angle $\theta$! Furthermore, in the model of [6,7,9] it is impossible to construct a beam totally polarized along $X$ or $y$ axes, as well as any elliptically polarized beam. Despite that each plane wave in the model of [6,7,9] has the same polarization in its own coordinate system, all of them are differently polarized in the beam reference system, which brings about a non-uniform polarization in the beam cross-section (see Fig. 3 below). It is worth noting that there is no optical device that is able to produce a beam with the polarization distribution (18) or (20). Indeed, when forming a polarized beam, the light passes through a polarizer oriented in a certain way with respect to common coordinate system $(X, y, Z)$. One cannot make a polarizer which would be oriented for each plane wave according to its own coordinate system $(X', Y', Z')$. For instance, if a linear polarizer is orientated along $X$ axis, so that it cuts $y$ component of the field, then the output beam will be polarized exactly as in Eq. (22), but not as in Eq. (20), since at $m_c = 0$ $\tilde{e}_y^{[II]} = 0$, but $\tilde{e}_y^{[I]} = \nu_y \cot\theta \neq 0$.

To determine the electric field for the whole incident beam, one has to set distribution function for the plane waves or for $\kappa$. We will consider a beam with Gaussian distribution, assuming for simplicity that the incident beam is cylindrically symmetric [15]. Then

$$\mathbf{E}(\mathbf{r}) \propto \iint d\kappa_X d\kappa_y \tilde{\mathbf{E}}(\mathbf{\kappa}) \exp\left[i\kappa_X X + i\kappa_y y + ik_Z Z\right], \quad \tilde{\mathbf{E}} = \tilde{\mathbf{e}} \exp\left[-\frac{\kappa_X^2 + \kappa_y^2}{2kD_0}\right]. \tag{23}$$

Here $D_0$ is a complex parameter which characterizes the width and phase front curvature of the beam at $Z = 0$ and $k_Z = \sqrt{k^2 - (\kappa_X^2 + \kappa_y^2)} \simeq k - (\kappa_X^2 + \kappa_y^2)/2k$ [16]. In equation (23) and in what follows we write the proportionality sign omitting inessential common amplitude factors. By introducing dimensionless integration variables in Eq. (23), $\kappa_i \to \nu_i = \kappa_i/k$, equation (23) can be given as

$$\mathbf{E} \propto \iint d\nu_X d\nu_y \tilde{\mathbf{e}}(\mathbf{v}) \exp\left[ik\tilde{\psi}(\mathbf{v})\right]. \tag{24}$$

Here $\tilde{\psi} = (\nu_X X + \nu_y y) + i(\nu_X^2 + \nu_y^2)/2D + Z$, whereas $D(Z) = D_0/(1 + iD_0 Z)$ characterizes the current width and the phase front curvature of the beam with diffraction in a homogeneous medium taken into account [17]. By evaluating integral (24) by the saddle-point method at $k \to \infty$, we obtain $\mathbf{E} \propto \mathbf{e} \exp(ik\psi)$. Here $\mathbf{e} = \mathbf{e}(\mathbf{r})$ and $\psi = \psi(\mathbf{r})$ are, respectively, the polarization vector and eikonal in the coordinate representation, at that $\mathbf{e}(\mathbf{r}) = \tilde{\mathbf{e}}(\mathbf{v}^*)$ and $\psi(\mathbf{r}) = \tilde{\psi}(\mathbf{v}^*)$, where $\mathbf{v}^*$ is determined from the condition $\partial\tilde{\psi}/\partial\mathbf{v} = 0$:

$$\nu_X^* = iDX, \quad \nu_y^* = iDy \tag{25}$$

As a result we have:

$$\mathbf{E} \propto \mathbf{e} \exp\left[-\frac{kD(X^2 + y^2)}{2} + ikZ\right], \tag{26}$$

where the two models of the beam polarization of [6,7,9] and [8] (from equations (20) and (22)), respectively, yield:

$$\mathbf{e}^{[I]} = \frac{(1 - im_c Dy \cot\theta)\mathbf{u}_X + (m_c + iDy \cot\theta)\mathbf{u}_y - i(DX + im_c Dy)\mathbf{u}_Z}{\sqrt{1 + |m_c|^2}}. \tag{27}$$

$$\mathbf{e}^{[II]} = \frac{\mathbf{u}_X + m_c \mathbf{u}_y - i(DX + m_c Dy)\mathbf{u}_Z}{\sqrt{1 + |m_c|^2}}. \tag{28}$$

Expressions (27) and (28) characterize the polarization distribution in the beam cross-section. In the second model, Eq. (28), the polarization is uniform and is characterized by the same parameter $m_c$: $m^{[II]} \equiv e_y^{[II]}/e_X^{[II]} = m_c$. The polarization of the first model, Eq. (27), as in $\kappa$-representation, is featured by some peculiarities. First, vector $\mathbf{e}^{[I]}$ turns out to be non-normalized:

$$\left|\mathbf{e}^{[I]}\right|^2 = 1 + \frac{4y\,\mathrm{Im}\,m_c\,\mathrm{Re}\,D\cot\theta}{1+|m_c|^2}. \tag{29}$$

Although the polarization vector was normalized in the wave vector representation at real $\mathbf{v}$, in the coordinate representation it is determined by a complex value of $\mathbf{v}$, Eq. (25). Second, the beam polarization $\mathbf{e}^{[I]}$ is non-uniform in the beam cross-section and is characterized by parameter $m^{[I]} = e_X^{[I]}/e_y^{[I]} = \tilde{m}^{[I]}(\mathbf{v}^*)$:

$$m^{[I]} = \frac{m_c + iDy\cot\theta}{1 - im_c Dy\cot\theta}. \tag{30}$$

In contrast to (21), because of complexity of $\mathbf{v}^*$, polarization $m_c$ gives rise to polarization $m^{[I]}$ with different eccentricity of the polarization ellipse. However, as before, circular polarization $m_c = \pm i$ correspond to the same circular polarization $m^{[I]} = \pm i$. This explains the fact that the approaches of [6,7,9] and [8] bring about the same results in the case of the circular polarization of the incident beam (see [8]). For linear polarization $m_c$, $\mathrm{Im}\,m_c = 0$, the polarization $m^{[I]}$ will be elliptical: $\mathrm{Im}\,m^{[I]} \neq 0$. For instance, if $m_c = 0$ or $m_c = \infty$, then $m^{[I]} = iDy\cot\theta$ or $m^{[I]} = i/Dy\cot\theta$. Figure 3 depicts an example of the polarization distribution in the cross-section of linearly-polarized beams of the two models at $m_c = 0$. One can see that the beam of the model of [6,7,9] is slightly depolarized: its edges are elliptically polarized with opposite helicities. Evidently, a real linear polarizer cannot generate such a beam. On the contrary, the beam of our model [8] is uniformly polarized along $X$ axis, as it should be in a real physical situation.

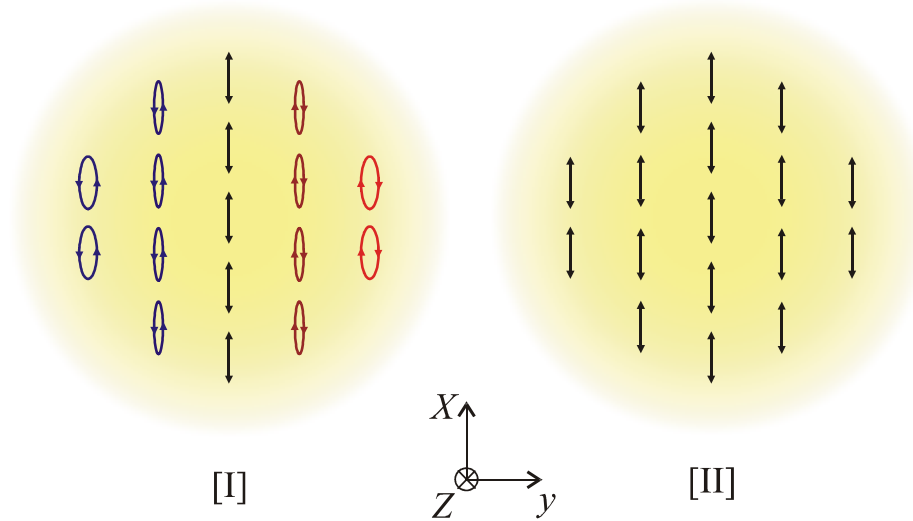

**Fig. 3.** (Color online). Scheme of the polarization distribution in the beam cross-section in the models [I] and [II]. A case of linearly polarized beam, $m_c = 0$, is presented.

The above argumentation is sufficient to give preference for the second beam model against the first one. Nevertheless, for completeness let us consider the Pointing and angular momentum vectors in the two beam models. By calculating the beam magnetic field $\mathbf{H} = -ik^{-1}\nabla\times\mathbf{E}$ (for simplicity we here assume $\varepsilon_1 = \mu_1 = 1$) with Eqs. (26)–(28), we arrive at the Pointing vector $\mathbf{S} \propto \mathrm{Re}(\mathbf{E}\times\mathbf{H}^*)$:

$$S_X^{[I]} = S_X^{[II]} \propto |F|^2 \left[ -X \operatorname{Im} D - \frac{2y \operatorname{Im} m_c \operatorname{Re} D}{1 + |m_c|^2} \right], \qquad (31a)$$

$$S_y^{[I]} = S_y^{[II]} \propto |F|^2 \left[ -y \operatorname{Im} D + \frac{2X \operatorname{Im} m_c \operatorname{Re} D}{1 + |m_c|^2} \right], \qquad (31b)$$

$$S_Z^{[I]} \propto |F|^2 \left[ 1 + \frac{4y \operatorname{Im} m_c \operatorname{Re} D \cot \theta}{1 + |m_c|^2} \right], \quad S_Z^{[II]} \propto |F|^2, \qquad (31c)$$

where $F = \exp\left[-kD(X^2 + y^2)/2\right]$. As seen from Eq. (31c), the longitudinal $Z$ component of the energy flux density is different in the two models. In the paraxial approximation under consideration (linear in $DX$ and $Dy$) the absolute value of the Pointing vector is determined by $Z$ component (31c) only and its difference in two models is directly related to the non-unit absolute value of the polarization vector in the first model, Eq. (29). The flux of the TAM in the beam is determined by the integral over the beam cross-section: $\mathbf{J} \propto \iint \mathbf{r} \times \mathbf{S}\, dXdy$. It can be easily shown that

$$J_Z^{[I]} = J_Z^{[II]} = J, \quad J_y^{[I]} = J_y^{[II]} = J_X^{[II]} = 0, \qquad (32a)$$

$$J_X^{[I]} \propto 4 \operatorname{Im} m_c \operatorname{Re} D \cot \theta \iint y^2 \exp\left[-k \operatorname{Re} D(X^2 + y^2)\right] dXdy = \frac{\pi \operatorname{Im} m_c \cot \theta}{k^2 \operatorname{Re} D}, \qquad (32b)$$

Thus, in the first beam model a non-zero transverse $X$ component of TAM is presented, which makes the angular momentum non-collinear to the energy flux: $\mathbf{J}^{[I]} \not\parallel \mathbf{S}^{[I]}$. As a consequence, the normal to the interface $z$ component of TAM is different in the first and the second models of the beam, which results in different values of TSs in two approaches. In addition, as is the case with the polarization vector (20), the Pointing vector (31) and angular momentum (32) of the incident beam in the first model depends, in natural beam coordinate system $(X, y, Z)$, upon the incidence angle $\theta$, which is meaningless from the physical viewpoint.

## IV. FIELDS OF REFLECTED AND REFRACTED BEAMS

Let us recur to the problem of the beam partial reflection at the interface between two media. The relations between the electric fields of the incident and scattered beams are determined by standard boundary conditions: continuity of tangent components of the electric and magnetic fields. As is known, for plane waves these conditions yield the Fresnel equations [1]. Therefore it is easier to solve the problem in the wave vector representation by applying the Fresnel equations to each plane wave in the incident packet.

Let $\tilde{\mathbf{E}}^a$ be the electric field of the respective plane wave with wave vector $\mathbf{k}^a$ (as before, $a = (i), (r), (t)$), and $(X'^a, Y'^a, Z'^a)$ is the coordinate system attached to $\mathbf{k}^a$ (similarly to $(X', Y', Z')$ coordinate system for the incident beam in the previous Section, Fig. 2). Then, the Fresnel equations read

$$\tilde{E}_{X'^a}^a = \tilde{\rho}_\parallel^a \tilde{E}_{X'}, \quad \tilde{E}_{Y'^a}^a = \tilde{\rho}_\perp^a \tilde{E}_{Y'}, \qquad (33)$$

where $\tilde{\rho}_\parallel^{(i,r,t)} = 1, R_\parallel, T_\parallel$ and $\tilde{\rho}_\perp^{(i,r,t)} = 1, R_\perp, T_\perp$ are the Fresnel coefficients for a current incident plane wave. The relation between orts of $(X'^a, Y'^a, Z'^a)$ coordinate system and $(X^a, Y^a, Z^a)$ coordinate system accompanying $a$ th beam is determined in paraxial approximation analogously to Eq. (18):

$$\mathbf{u}_{X'^a} = \mathbf{u}_{X^a} + v_y^a \cot \vartheta^a \mathbf{u}_y - v_{X^a}^a \mathbf{u}_{Z^a},$$

$$\mathbf{u}_{Y'^a} = \mathbf{u}_y - \nu_y^a \cot \vartheta^a \mathbf{u}_{X^a} - \nu_y^a \mathbf{u}_{Z^a}, \tag{34}$$
$$\mathbf{u}_{Z'^a} = \mathbf{u}_{Z^a} + \nu_{X^a}^a \mathbf{u}_{X^a} + \nu_y^a \mathbf{u}_y,$$

where, as in the previous Section, the wave vector of a plane wave from $a$ th beam is represented as $\mathbf{k}^a = \mathbf{k}_c^a + \mathbf{\kappa}^a$ and $\mathbf{v}^a = \mathbf{\kappa}^a / k^a$ ($k^a = k_c^a$). From the Snell law it follows in paraxial approximation that

$$\nu_y^a = \frac{k}{k^a} \nu_y, \quad \nu_{X^a}^a = \frac{k \cos \theta}{k^a \cos \vartheta^a} \nu_X. \tag{35}$$

By using representation $\tilde{\mathbf{E}}^a = \tilde{E}_{X'^a}^a \mathbf{u}_{X'^a} + \tilde{E}_{Y'^a}^a \mathbf{u}_{Y'^a}$, equations (33)–(35), and connection $\tilde{E}_{X'} = \tilde{E}_X + \nu_y \cot \theta \tilde{E}_y$, $\tilde{E}_{Y'} = \tilde{E}_y - \nu_y \cot \theta \tilde{E}_X$ (which follows from Eq. (19) and $\tilde{E}_Z \sim \nu$), we obtain the expressions for components of the $a$ th beam in the respective beam coordinate system $(X^a, y, Z^a)$:

$$\tilde{\mathbf{E}}^a = \tilde{E}_{X^a}^a \mathbf{u}_{X^a} + \tilde{E}_y^a \mathbf{u}_y + \tilde{E}_{Z^a}^a \mathbf{u}_{Z^a}, \tag{36a}$$

$$\tilde{E}_{X^a}^a = \tilde{\rho}_\parallel^a \tilde{E}_X + \frac{\nu_y^a}{\sin \vartheta^a} \left( \tilde{\rho}_\parallel^a \cos \theta - \tilde{\rho}_\perp^a \cos \vartheta^a \right) \tilde{E}_y, \tag{36b}$$

$$\tilde{E}_y^a = \frac{\nu_y^a}{\sin \vartheta^a} \left( \tilde{\rho}_\parallel^a \cos \vartheta^a - \tilde{\rho}_\perp^a \cos \theta \right) \tilde{E}_X + \tilde{\rho}_\perp^a \tilde{E}_y, \tag{36c}$$

$$\tilde{E}_{Z^a}^a = -\left( \nu_{X^a}^a \tilde{\rho}_\parallel^a \tilde{E}_X + \nu_y^a \tilde{\rho}_\perp^a \tilde{E}_y \right). \tag{36d}$$

The Fresnel coefficients depend on the angle of incidence of a given plane wave and, therefore, $\tilde{\rho}_\parallel^a = \tilde{\rho}_\parallel^a(\mathbf{v})$, $\tilde{\rho}_\perp^a = \tilde{\rho}_\perp^a(\mathbf{v})$. In the linear approximation in $\mathbf{v}$ one can put

$$\tilde{\rho}_{\parallel,\perp}^a = \rho_{\parallel,\perp c}^a + \left. \frac{d\tilde{\rho}_{\parallel,\perp}^a}{d\nu_X} \right|_{\mathbf{v}=0} \nu_X, \tag{37}$$

where $\rho_{\parallel,\perp c}^a = \tilde{\rho}_{\parallel,\perp}^a(0)$ and we have taken into account that expansion of $\tilde{\rho}_{\parallel,\perp}^a$ in terms of $\nu_y$ starts with square terms. The correction proportional to $\nu_X$ in Eq. (37) is responsible for the effect of *longitudinal shift* of the beam (the Goos–Hänchen shift) [18] and, as we will see below, it does not contribute to the Fedorov–Imbert transverse shift to be considered.

Substitution of Eqs. (20), (22), and (23) with Eq. (35) in Eq. (36) allows to represent the electric field of $a$ th beam as:

$$\tilde{\mathbf{E}}^a = \frac{\tilde{\rho}_\parallel^a}{\rho_{\parallel c}^a} A_c^a \exp\left[ -k^a \left( \frac{\nu_{X^a}^{a2}}{2D_{0X^a}^a} + \frac{\nu_y^{a2}}{2D_{0y}^a} \right) \right] \tilde{\mathbf{e}}^a, \tag{38}$$

where $A_c^a = \sqrt{\rho_{\parallel c}^{a2} + \rho_{\perp c}^{a2} |m_c|^2} / \sqrt{1 + |m_c|^2} \operatorname{sgn} \rho_{\parallel c}^a$ or $A_c^{(i,r,t)} = 1, R_c \operatorname{sgn} R_{\parallel c}, T_c$ are the amplitude scattering coefficients for the central plane wave, Eqs. (14) and (15), $D_{0X^a}^a = \frac{k \cos^2 \theta}{k^a \cos^2 \vartheta^a} D_0$ and $D_{0y}^a = \frac{k}{k^a} D_0$ are the parameters characterizing the widths and the phase front curvatures of $a$ th beam along $X^a$ and $y$ axes at $Z^a = 0$ [19], and $\tilde{\mathbf{e}}^a$ is the polarization vector of the current plane wave in $a$ th beam [20]. Two models of the incident beam, Eqs. (20) and (22), yield, respectively:

$$\tilde{\mathbf{e}}^{a[I]} = \frac{1}{\sqrt{1 + |m_c^a|^2}} \left\{ \left[ 1 - \tilde{m}^a \nu_y^a \cot \vartheta^a \right] \mathbf{u}_{X^a} + \left[ \tilde{m}^a + \nu_y^a \cot \vartheta^a \right] \mathbf{u}_y - \left[ \nu_{X^a}^a + \tilde{m}^a \nu_y^a \right] \mathbf{u}_{Z^a} \right\}, \tag{39}$$

$$\tilde{\mathbf{e}}^{a[\text{II}]} = \frac{1}{\sqrt{1+\left|m_{\text{c}}^{a}\right|^{2}}}\left\{\left[1-\tilde{m}^{a}v_{y}^{a}\cot\vartheta^{a}\left(1-\frac{\cos\theta}{\tilde{\rho}^{a}\cos\vartheta^{a}}\right)\right]\mathbf{u}_{X^{a}}\right.$$
$$\left.+\left[\tilde{m}^{a}+v_{y}^{a}\cot\vartheta^{a}\left(1-\frac{\tilde{\rho}^{a}\cos\theta}{\cos\vartheta^{a}}\right)\right]\mathbf{u}_{y}-\left[v_{X^{a}}^{a}+\tilde{m}^{a}v_{y}^{a}\right]\mathbf{u}_{Z^{a}}\right\}. \qquad (40)$$

Here $\tilde{m}^{a} = \tilde{\rho}^{a}m_{\text{c}}$, $\tilde{\rho}^{a} = \tilde{\rho}_{\perp}^{a}/\tilde{\rho}_{\parallel}^{a}$, whereas $m_{\text{c}}^{a} = \rho_{\text{c}}^{a}m_{\text{c}}$ ($\rho_{\text{c}}^{a} = \rho_{\perp\text{c}}^{a}/\rho_{\parallel\text{c}}^{a}$) is the parameter of polarization of the central plane wave of $a$ th beam, Eq. (14). For the central plane wave, $\mathbf{v} = 0$, both polarizations (39) and (40) equals

$$\mathbf{e}_{\text{c}}^{a} = \tilde{\mathbf{e}}_{\text{c}}^{a} = \tilde{\mathbf{e}}^{a}(0) = \frac{\mathbf{u}_{X^{a}} + m_{\text{c}}^{a}\mathbf{u}_{y}}{\sqrt{1+\left|m_{\text{c}}^{a}\right|^{2}}}. \qquad (41)$$

Fields (38)–(40) can be converted into coordinate representation similarly to Eqs. (23)–(26):

$$\mathbf{E}^{a}(\mathbf{r}) \propto \iint d\kappa_{X^{a}}^{a}d\kappa_{y}^{a}\tilde{\mathbf{E}}^{a}(\mathbf{\kappa}^{a})\exp\left[i\kappa_{X^{a}}^{a}X^{a}+i\kappa_{y}^{a}y+ik_{Z^{a}}^{a}Z^{a}\right]. \qquad (42)$$

By substituting Eq. (38) and $k_{Z^{a}}^{a} \simeq k^{a} - \left(\kappa_{X^{a}}^{a2}+\kappa_{y}^{a2}\right)/2k^{a}$, integral (42) can be represented as

$$\mathbf{E}^{a} \propto A_{\text{c}}^{a}\iint dv_{X^{a}}^{a}dv_{y}^{a}\frac{\tilde{\rho}_{\parallel}^{a}}{\rho_{\parallel\text{c}}^{a}}\tilde{\mathbf{e}}^{a}(\mathbf{v}^{a})\exp\left[ik^{a}\tilde{\psi}^{a}(\mathbf{v}^{a})\right]. \qquad (43)$$

Here

$$\tilde{\psi}^{a} = v_{X^{a}}^{a}X^{a}+v_{y}^{a}y+i\left(\frac{v_{X^{a}}^{a2}}{2D_{X^{a}}^{a}}+\frac{v_{y}^{a2}}{2D_{y}^{a}}\right)+Z^{a}, \qquad (44)$$

whereas parameters

$$D_{X^{a}}^{a}(Z^{a}) = \frac{D_{0X^{a}}^{a}}{1+iD_{0X^{a}}^{a}Z^{a}}, \quad D_{y}^{a}(Z^{a}) = \frac{D_{0y}^{a}}{1+iD_{0y}^{a}Z^{a}} \qquad (45)$$

characterize the current widths and the phase front curvatures of the beam along $X^{a}$ and $y$ axes with the diffraction phenomena taken into account [17]. Evaluating integral (43) by the saddle-point method at $k^{a} \to \infty$, we obtain $\mathbf{E}^{a} \propto \frac{\rho_{\parallel}^{a}}{\rho_{\parallel\text{c}}^{a}}A_{\text{c}}^{a}\exp(ik\psi^{a})\mathbf{e}^{a}$. Here $\mathbf{e}^{a}(\mathbf{r})$ and $\psi^{a}(\mathbf{r})$ are, respectively, the polarization vector and eikonal in the coordinate representation, with $\mathbf{e}^{a}(\mathbf{r}) = \tilde{\mathbf{e}}^{a}(\mathbf{v}^{a*})$ and $\psi^{a}(\mathbf{r}) = \tilde{\psi}^{a}(\mathbf{v}^{a*})$, where $\mathbf{v}^{a*}$ is determined from $\partial\tilde{\psi}^{a}/\partial\mathbf{v} = 0$:

$$v_{X^{a}}^{a*} = iD_{X^{a}}^{a}X^{a}, \quad v_{y}^{a*} = iD_{y}^{a}y. \qquad (46)$$

Coordinate representations for other quantities are defined similarly: $\rho_{\parallel,\perp}^{a}(\mathbf{r}) = \tilde{\rho}_{\parallel,\perp}^{a}(\mathbf{v}^{a*})$, $m^{a}(\mathbf{r}) = \tilde{m}^{a}(\mathbf{v}^{a*}) = \rho^{a}m_{\text{c}}$, and $\rho^{a}(\mathbf{r}) = \tilde{\rho}^{a}(\mathbf{v}^{a*}) = \rho_{\perp}^{a}/\rho_{\parallel}^{a}$. In so doing, from Eqs. (37) and (46) it follows that

$$\rho_{\parallel,\perp}^{a}(\mathbf{r}) = \rho_{\parallel,\perp\text{c}}^{a}+i\left.\frac{d\tilde{\rho}_{\parallel,\perp}^{a}}{dv_{X}}\right|_{\mathbf{v}=0}D_{X^{a}}^{a}X^{a}. \qquad (47)$$

As a result we have:

$$\mathbf{E}^{a} \propto \frac{\rho_{\parallel}^{a}}{\rho_{\parallel\text{c}}^{a}}A_{\text{c}}^{a}\mathbf{e}^{a}\exp\left[-\frac{k^{a}\left(D_{X^{a}}^{a}X^{a2}+D_{y}^{a}y^{2}\right)}{2}+ik^{a}Z^{a}\right], \qquad (48)$$

where the two models of the beam polarization (from equations (39) and (40)), respectively, yield:

$$\mathbf{e}^{a[\mathrm{I}]} = \frac{1}{\sqrt{1+|m^a|^2}} \left\{ \left[1 - im^a D_y^a y \cot \vartheta^a \right] \mathbf{u}_{X^a} \right.$$
$$\left. + \left[ m^a + i D_y^a y \cot \vartheta^a \right] \mathbf{u}_y - i \left[ D_{X^a}^a X^a + m^a D_y^a y \right] \mathbf{u}_{Z^a} \right\}, \tag{49}$$

$$\mathbf{e}^{a[\mathrm{II}]} = \frac{1}{\sqrt{1+|m^a|^2}} \left\{ \left[1 - im^a D_y^a y \cot \vartheta^a \left(1 - \frac{\cos\theta}{\rho^a \cos\vartheta^a}\right) \right] \mathbf{u}_{X^a} \right.$$
$$\left. + \left[ m^a + i D_y^a y \cot \vartheta^a \left(1 - \frac{\rho^a \cos\theta}{\cos\vartheta^a}\right) \right] \mathbf{u}_y - i \left[ D_{X^a}^a X^a + m^a D_y^a y \right] \mathbf{u}_{Z^a} \right\}. \tag{50}$$

Equations (38)–(40) and (48)–(50) describe wave vector and coordinate representations of the electric fields of the incident, reflected and refracted beams, thereby solving the problem of the Gaussian beam scattering at the interface in the two models. At $a = (i)$, as it should be, they are converted to Eqs. (20), (22), (23) and (22)–(26) introduced for the incident beam. As it is shown in [3,8], the reflected and refracted beams in both models do not represent pure polarization states anymore. In particular, for linearly-polarized incident beam, Eq. (22) with $m_c = 0$, Fig. 3b, the scattered beams will have a non-uniform polarization distribution resembling that in Fig. 3a.

## V. TRANSVERSE SHIFTS

To determine TS of the center of gravity of $a$ th beam we introduce field (48)–(50) integrated over $X^a$ coordinate: $\bar{\mathbf{E}}^a \propto \int \mathbf{E}^a dX^a$ [21]; it enables one not to take the longitudinal shift into account. Under integration, linear in $X^a$ terms vanish (higher-order terms should be neglected) and for the first and second beam models, Eqs. (49) and (50), we have:

$$\bar{\mathbf{E}}^{a[\mathrm{I}]} \propto \frac{A_c^a}{\sqrt{1+|m_c^a|^2}} \exp\left[-\frac{k^a D_y^a y^2}{2} + ik^a Z^a \right]$$
$$\times \left\{ \left[1 - im_c^a D_y^a y \cot \vartheta^a \right] \mathbf{u}_{X^a} + \left[ m_c^a + i D_y^a y \cot \vartheta^a \right] \mathbf{u}_y - im_c^a D_y^a y \mathbf{u}_{Z^a} \right\}, \tag{51}$$

$$\bar{\mathbf{E}}^{a[\mathrm{II}]} \propto \frac{A_c^a}{\sqrt{1+|m_c^a|^2}} \exp\left[-\frac{k^a D_y^a y^2}{2} + ik^a Z^a \right]$$
$$\times \left\{ \left[1 - im_c^a D_y^a y \cot \vartheta^a \left(1 - \frac{\cos\theta}{\rho_c^a \cos\vartheta^a}\right) \right] \mathbf{u}_{X^a} \right.$$
$$\left. + \left[ m_c^a + i D_y^a y \cot \vartheta^a \left(1 - \frac{\rho_c^a \cos\theta}{\cos\vartheta^a}\right) \right] \mathbf{u}_y - im_c^a D_y^a y \mathbf{u}_{Z^a} \right\}, \tag{52}$$

Note that expression (52) is equivalent to the formulas for the electric fields of beams unconfined along $X^a$, which have been derived in paper [8] (equation (4) therein).

The transverse $y$ coordinate of the center of gravity of $a$ th beam, $y_c^a$, is defined as the position of the maximum of function $f(y) = |\bar{\mathbf{E}}^a|^2$. For a paraxial Gaussian beam such a definition is equivalent to the projection of the field onto the central polarization vector (41), which have been used in [6,8]. In the linear approximation in $D_y^a y$ the square of the absolute value of fields (51) and (52) can be written as

$$\left|\overline{\mathbf{E}}^{a[\mathrm{I},\mathrm{II}]}\right|^{2} \propto A_{c}^{a2} \exp\left[-k^{a} \operatorname{Re} D_{y}^{a} \left(y - y_{c}^{a[\mathrm{I},\mathrm{II}]}\right)^{2}\right], \tag{53}$$

where, in two models, $y_c^a$ equals

$$y_{c}^{a[\mathrm{I}]} = \frac{2 \operatorname{Im} m_{c}^{a} \cot \vartheta^{a}}{k^{a}\left(1 + \left|m_{c}^{a}\right|^{2}\right)}, \tag{54}$$

$$y_{c}^{a[\mathrm{II}]} = \frac{\cot \vartheta^{a}}{k^{a}\left(1 + \left|m_{c}^{a}\right|^{2}\right)} \left\{ \operatorname{Im} m_{c}^{a}\left[2 - \left(\rho_{c}^{a} + \rho_{c}^{a-1}\right)\frac{\cos\theta}{\cos\vartheta^{a}}\right] + \operatorname{Re} m_{c}^{a} \frac{\operatorname{Im} D_{y}^{a}}{\operatorname{Re} D_{y}^{a}} \left(\rho_{c}^{a} - \rho_{c}^{a-1}\right)\frac{\cos\theta}{\cos\vartheta^{a}} \right\}. \tag{55}$$

It is worth noting that $y_{c}^{(i)[\mathrm{II}]} = 0$, whereas $y_{c}^{(i)[\mathrm{I}]} = \frac{2 \operatorname{Im} m_{c} \cot \theta}{k\left(1 + \left|m_{c}\right|^{2}\right)} \neq 0$. That means that the center of the incident beam of the model of [6,7,9] is shifted from $y = 0$ plane (despite the fact that the initial Gaussian distribution had a maximum at $y = 0$). The shift $y_{c}^{(i)[\mathrm{I}]}$ originates from the form of polarization vector (20) in the beam coordinate system. Indeed, it has $v_y$-dependent coefficients at $\mathbf{u}_X$ and $\mathbf{u}_y$, which implies that each plane wave in the incident beam has its own $v_y$-dependent phase that shifts the center of gravity of the beam.

TS of $a$ th beam (relative to the incident beam) equals $\delta y^a = y_c^a - y_c^{(i)}$. It can be given as a sum of two terms, $\delta y^a = \Delta^a + \delta^a$, where

$$\Delta^{a[\mathrm{I}]} = \frac{2 \operatorname{Im} m_{c} \cot \theta}{k}\left[\frac{\rho_{c}^{a} \cos \vartheta^{a} / \cos \theta}{1 + \rho_{c}^{a2}\left|m_{c}\right|^{2}} - \frac{1}{1 + \left|m_{c}\right|^{2}}\right], \tag{56}$$

$$\delta^{a[\mathrm{I}]} = 0, \tag{57}$$

$$\Delta^{a[\mathrm{II}]} = \frac{\operatorname{Im} m_{c} \cot \theta}{k} \frac{2\rho_{c}^{a} \cos \vartheta^{a} / \cos \theta - 1 - \rho_{c}^{a2}}{1 + \rho_{c}^{a2}\left|m_{c}\right|^{2}}, \tag{58}$$

$$\delta^{a[\mathrm{II}]} = \frac{\operatorname{Re} m_{c} \operatorname{Im} D_{y}^{a} \cot \theta}{k \operatorname{Re} D_{y}^{a}} \frac{\rho_{c}^{a2} - 1}{1 + \rho_{c}^{a2}\left|m_{c}\right|^{2}}. \tag{59}$$

Here we endeavored to express TSs in terms of parameters of the incident beam and Fresnel coefficients. TSs $\Delta^a$ are the desired Fedorov–Imbert shift. Equation (58) coincides with the analogous equation (5) in paper [8], but differs from equation (56), which represents the TS obtained in papers [6,7,9]. TS $\Delta^{a[\mathrm{I}]}$, Eq. (56), coincides with Eq. (17) and, hence, fulfil one-photon TAM conservation laws (11) as well as general TAM conservation law (12) and (16). TS $\Delta^{a[\mathrm{II}]}$, Eq. (58) also satisfies the general conservation law (12) and (16), but in general case does not fulfill one-photon conservation laws (11) (with Eqs. (10) and (13)). The results of approaches of [6,7,9] and [8] coincide, $\Delta^{a[\mathrm{II}]} = \Delta^{a[\mathrm{I}]}$, only when $\left|m_c\right| = 1$ (e.g., for circularly polarized incident wave, $m_c = \pm i$) and in the trivial case of linear polarization, $\operatorname{Im} m_c = 0$, $\Delta^a = 0$, see also [22]. In other cases, the difference between shifts (56) and (58) is of the order of the TS itself (a fraction of the wavelength) and, hence, can be detected with an accuracy of nowaday experiments [2]. Remarkably, Eqs. (56) and (58) follow from the general expressions of [5] when substituting the incident beam field from Eq. (20) or (22), respectively.

The second summand in the transverse shift of the second model, $\delta^{a[\mathrm{II}]}$, can be regarded as a small *angular transverse shift* inclining the plane of the beam propagation with respect to $y = const$ plane. Indeed, assume that $\operatorname{Im} D_{0y}^{a} = 0$, which correspond to the situation where the beam focus is located at $Z^a = 0$. Then from Eq. (45) it follows that $\dfrac{\operatorname{Im} D_y^a}{\operatorname{Re} D_y^a} = -D_{0y}^a Z^a$ and shift $\delta^{a[\text{II}]}$ grows linearly with $Z^a$. Taking into account that $D_{0y}^a \sim \lambda / w^2$ (where $\lambda$ is the wavelength and $w$ is the beam width in the focus), the angular shift $\delta^{a[\text{II}]}$ can be estimated as $\delta^{a[\text{II}]} \sim \lambda^2 Z^a / w^2$, while the linear shift $\Delta^a \sim \lambda$. Both shifts become of the same order at the distance of the order of the Rayleigh range, $Z^a \sim w^2 / \lambda$. The angular shift vanishes in the model of [6,7,9], Eq. (57). At the same time, the possibility of such shift has been pointed out in [5], Eq. (19) therein, but neglected in subsequent calculations due to diffractionless approximation used. Concrete expressions of the angular shift have been first calculated by Nasalski [2] for the reflected beam. The linear (Fedorov–Imbert) shift is a geometrical optics effect, whereas the angular shift is a diffraction phenomenon quadratic in the wavelength. That is why the latter one cannot be accounted for by simple particle (or geometrical optics) considerations and respective conservation laws. Note that $\delta^{a[\text{II}]} \neq 0$ only at $\operatorname{Re} m \neq 0$, i.e. when the polarization ellipse is inclined with respect to $y$ axis. This gives an additional asymmetry of the field relative to $y = 0$ plane, which can cause the angular shift. As far as we know, the angular transverse shift, Eq. (59), for the general case of partial reflection and refraction is calculated here explicitly for the first time.

### VI. CONCLUSION

We have examined the problem of partial reflection and refraction of a paraxial Gaussian electromagnetic beam. We paid a special attention to the transverse shift of the centers of gravity of the reflected and refracted beams as well as to the conservation laws for angular momentum of the field. It has been shown that the discrepancy between the results of previous papers [6,7,9] and [8] occurs not due to calculation errors, but owing to essentially different models of the polarization structure of the incident beam. Simple, on the first sight, model of [6,7,9] describes an unnatural beam which cannot be produced by real polarizers. First, the electric field of the incident beam in the coordinate frame attendant to its center is dependent on the incidence angle. Second, the polarization distribution in the plane orthogonal to the beam axis is non-uniform. Such a beam is, in fact, slightly depolarized. Finally, the angular momentum vector of the beam is inclined with respect to the beam axis. At the same time, a model of the beam we have proposed in [8] is free of all these drawbacks and fits for real physical beam passed through a respective polarizer.

Detailed calculations of the partial reflection and refraction of Gaussian beam confirmed the expressions for the transverse shifts obtained in [8]. These shifts satisfy the conservation law of the normal component of the total angular momentum in the problem. However, they do not obey, in general case, one-photon conservation laws suggested in [7,9] (see also [8]). It is worth noting that the difference between the results of papers [6,7,9] and [8] reveals itself only for elliptically polarized incident beam and in the process of multi-channel scattering. In the case of circularly polarized incident beam or at the total internal reflection both models result in the same transverse shifts.

In addition to linear Fedorov–Imbert transverse shifts, we have calculated transverse angular shifts of the reflected and refracted beams, which are related to the beams diffraction. These angular shifts are absent in the model of [6,7,9].

*Note added.* – After submission of this work, a comment [22] appears that confirms our conclusions. It is shown there that TSs both from papers [6,7,9] and from paper [8] are in agreement with the general expressions of [5], but differ from each other due to the difference in the incident beam fields. It is also noticed there that a general-form wave packet does not satisfy one-photon TAM conservation laws.

### ACKNOWLEDGMENTS

We are indebted to V. Fedoseev for his valuable comments and to H. Sasada, S. Murakami, and M. Onoda for fruitful correspondence. This work was partially supported by STCU (grant P-307) and CRDF (grant UAM2-1672-KK-06).